\documentstyle[psfig,amssymb]{mn}
\title[The XMM-Newton/2dF survey - II.  The nature of X-ray faint optically bright X-ray sources.]
{The XMM-Newton/2dF survey - II.  The nature of X-ray faint optically bright X-ray sources.}
 
\author[A. Georgakakis et al.] {A. Georgakakis$^{1}$\thanks{email:
  age@astro.noa.gr}, 
  I. Georgantopoulos$^{1}$,  M. Vallb\'e$^2$, V. Kolokotronis$^{1}$,
  S. Basilakos$^{1}$,
  \\ \\
  {\LARGE M. Plionis$^{1,3}$, G. C. Stewart$^4$, T. Shanks$^2$, B. J. Boyle$^5$} \\ \\  
  $^1$ Institute of Astronomy \& Astrophysics, National Observatory of
  Athens, I. Metaxa \& V. Pavlou, Athens, 15236, Greece \\ 
  $^2$ Physics Department, University of Durham,  Science Labs,
  South Road, Durham, DH1 3LE, UK \\
  $^3$ Instituto Nacional de Astrofisica, Optica y Electronica (INAOE), Apartado Postal 51 y 216, 72000, Puebla, Mexico\\
  $^4$ Department of Physics and Astronomy, University of Leicester
  Leicester LE1 7RH, UK \\
  $^5$ Anglo-Australian Observatory, PO Box 296, Epping, NSW 2121, Australia \\
}
\begin{document}
\maketitle  

\begin{abstract}
 In this paper we investigate the properties of low X-ray--to--optical
 flux ratio sources detected in a wide area ($2.5 \rm deg^2$) shallow
 [$f_X(\rm 0.5 - 8 \,keV) \approx 10^{-14} \,erg \,s^{-1} \,cm^{-2}$]
 XMM-{\it Newton} survey. We find a total of  26 sources ($5\%$ of the
 total X-ray selected population) with  $\log f_X/f_{opt} < -0.9$ to
 the above flux limit.  Optical spectroscopy is available for 20 of
 these low X-ray--to--optical flux ratio objects. Most of them are
 found to be associated with Galactic stars (total of 8) and broad
 line AGNs (total of 8).  We also find two sources with optical
 spectra showing absorption and/or narrow emission lines 
 and X-ray/optical properties suggesting AGN activity. 
 Another two sources are found to be associated with low redshift
 galaxies with narrow emission line optical spectra, X-ray
 luminosities $L_X(\rm 0.5-8\,keV)\rm \approx 
 10^{41}\,erg\,s^{-1}$ and $\log f_X/f_{opt} \approx -2$   
 suggesting `normal' star-forming galaxies. Despite the
 small number statistics the sky density of `normal' X-ray selected
 star-forming galaxies 
 at the flux limit of the present sample is low consistent with
 previous ROSAT HRI deep surveys. Also, the number density estimated
 here is in good agreement with both the $\log N -\log S$ of `normal'
 galaxies in the Chandra Deep Field North (extrapolated to bright
 fluxes) and model predictions based on the X-ray luminosity function
 of local star-forming galaxies.
\end{abstract}


\section{Introduction}\label{sec_intro}

X-ray surveys with the {\it ROSAT} and more recently with the {\it
Chandra} and  the XMM-{\it Newton} observatories have demonstrated
that the X-ray source  population is a heterogeneous mix of objects
comprising (i) powerful AGNs/QSOs, (ii) galaxy groups/clusters, (iii) 
low luminosity AGNs, (iv) `normal' galaxies with X-ray emission
dominated by stellar processes and (v) a small number of Galactic
stars (Lehmann et al. 2001; Barger et al. 2002; Hornschemeier et
al. 2003).   

Powerful AGNs are identified in X-ray surveys by their enhanced
X-ray--to--optical flux ratios   $\log f_X/f_{opt} \ga -1$. These  
sources are responsible for the bulk of the diffuse X-Ray Background
(XRB) and are therefore important for constraining models on 
the XRB origin. 

In addition to powerful AGNs, deep surveys with the {\it Chandra}
observatory reaching fluxes well below  $f(\rm 0.5 - 2 \,keV) \la
10^{-15} \,erg \,s^{-1} \,cm^{-2}$ have revealed large numbers of low 
X-ray--to--optical flux ratio sources $\log  f_X/f_{opt} \la
-1$. These objects although too faint to significantly contribute to
the XRB are detected in increasing numbers with decreasing flux and are
likely to outnumber  powerful AGNs below  $f(\rm 0.5 - 2 \,keV) \approx
10^{-17} \, erg\, s^{-1}\, cm^{-2}$ (Hornschemeier et al. 2003). The
low X-ray--to--optical flux ratio regime is thought to be populated by 
Low Luminosity AGNs (LLAGNs) and `normal' galaxies (Hornschemeier et
al. 2003). Indeed, sources with $-2\la \log f_X/f_{opt}\la   -1$
exhibit enhanced activity attributed to either starbursts or LLAGNs
(Alexander et al. 2002; Bauer et al. 2002;  Hornschemeier et
al. 2003; Georgakakis et al. 2003b). Even lower X-ray--to--optical
flux ratios,  $\log  f_X/f_{opt}  \la -2$, are  believed to be
quiescent Milky  Way type galaxies (Hornschemeier et al. 2002;
Georgakakis et al. 2003a; Hornschemeier et al. 2003) although the
presence of heavily  obscured AGNs cannot be excluded.

Deep Chandra surveys have provided valuable information on 
the nature of low $f_X/f_{opt}$ sources at fluxes $f(\rm
0.5 - 2 \,keV) \la 10^{-17} - 10^{-16} \,erg \,s^{-1} \,cm^{-2}$. 
These surveys are not however suitable for the study of low
$f_X/f_{opt}$ sources at brighter fluxes  because of the {\it Chandra}
small field-of-view ($\rm \approx0.07\,deg^{2}$). Indeed, due to the
low surface density of these sources at bright fluxes wide area
surveys are required to compile large statistically complete samples
to elucidate their nature.    

The {\it ROSAT}  satellite with a field-of-view of $\rm
\approx0.30\,deg^{2}$ per pointing has probed the bright flux regime
[$f(\rm 0.5 - 2 \,keV)  \approx  10^{-15}\,erg \,s^{-1} \,cm^{-2}$] but
has identified only a small number of low X-ray--to--optical flux
ratio sources (Griffiths et  al. 1995; Griffiths et al. 1996; 
Georgantopoulos et al. 1996; Lehmann et al. 2001). This can be
attributed to (i) the poor positional accuracy of ROSAT rendering the
optical  identification of X-ray sources difficult and (ii) its low
sensitivity making wide area surveys to the limit $f(\rm 0.5 - 2 \,keV)
\approx 10^{-15}\,erg  \,s^{-1} \,cm^{-2}$ time  consuming. 
The latter is particularly true due to the low surface density of
$f_X/f_{opt}\la -1$ sources to the flux limit above requiring large
surveyed regions to compile statistically complete samples.   
Moreover, the insensitivity of ROSAT to hard X-rays ($>2.5\rm\,keV$)
is a major drawback since heavily obscured AGNs (believed to populate
the $\log f_X/f_{opt}\la   -1$ regime) are expected to emit most of
their X-ray energy output in this spectral band.    

In this paper we employ a shallow  (2-10\,ks) wide area ($\rm
\approx 2.5\,deg^{2}$) XMM-{\it Newton} survey to explore the nature of the
$\log f_X / f_{opt} \la -1$  sources to the limit $f(\rm 0.5 - 8
\,keV) \approx 10^{-14}\,erg \,s^{-1} \,cm^{-2}$ similar to
that probed by previous {\it ROSAT} surveys. Compared to {\it ROSAT},
XMM-{\it Newton} has the advantage of significantly higher sensitivity
over a wide energy range (0.2-10\,keV) and improved positional accuracy
($\la3.5$\,arcsec; Hasinger et al. 2001; McHardy et al. 2003)
facilitating the optical identification of X-ray sources. Our XMM-{\it
Newton} observations (hereafter referred to as the XMM/2dF survey)
cover an area of $\rm \approx 2.5\,deg^{2}$ much  larger than
any previous {\it ROSAT} surveys at the same flux  limit. This is
essential to compile a large sample of low X-ray--to--optical flux
ratio sources to explore their nature and to constrain the relative
fraction of the different X-ray  populations. 
An advantage of the data presented here is that unlike deep
Chandra samples (e.g. Brandt et al. 2001) limited by the small 
field--of--view,  the XMM/2dF survey has sufficient areal 
coverage to probe the low-$z$ ($z\la0.1$)  universe. This is essential
to explore the nature of the relatively nearby X-ray population and
to provide a link between local ($\rm <100\,Mpc$) and more distant
samples.  

Moreover, our XMM/2dF survey has the advantage of high quality
follow-up optical spectroscopic and photometric observations. Apart
from our own spectroscopic campaign the  XMM/2dF overlaps with large
scale spectroscopic programs: the Sloan Digital Sky Survey (SDSS), the
2dF Galaxy and QSO  Redshift Surveys (2dFGRS and 2QZ
respectively). These unprecedented spectroscopic databases are
complemented by a homogeneous set of multi-waveband ($ugriz$)
photometric data from  the SDSS available for part of our XMM/2dF
survey.      

Section \ref{sec_survey} presents the XMM/2dF survey, section
\ref{sec_xray} describes the reduction of the X-ray data while sections
\ref{sec_phot} and \ref{sec_spec} detail the optical photometric and
spectroscopic observations respectively. In section \ref{sec_id} we
outline the optical identification method while section \ref{sample}
presents the sample employed in this study. The results are discussed
in section \ref{discussion}.  Finally section \ref{conclusions}
summarises  our conclusions.  Throughout this paper we adopt
$\rm H_{o}=65\,km\,s^{-1}\,Mpc^{-1}$, $\rm \Omega_{M}=0.3$ and
$\rm \Omega_{\Lambda}=0.7$.   

\section{The XMM/2dF Survey}\label{sec_survey}
The North Galactic Pole F864 region [RA(J2000)=$13^{\rm h}41^{\rm m}$; 
Dec.(J2000)=$00\degr00\arcmin$] and the South Galactic Pole
[SGP; RA(J2000)=$00^{\rm h} 57^{\rm m}$, Dec.(J2000)=$-28\degr
00\arcmin$] were surveyed by the XMM-{\it Newton} between May 2002 and 
February 2003 as part of the Guaranteed Time program. The observations
consist of a total of 18 pointings split between the SGP (total of 5)
and the F864 (total of 8) regions each  with an exposure time of
$\approx2-10$\,ks. Part of this dataset was presented by Georgakakis 
et al. (2003a). In the present paper the Georgakakis et al. (2003a)
observations are  supplemented by a total of three additional
pointings recently  obtained by  the XMM-{\it Newton} in the F864
region. These three fields suffered from elevated particle background
or instrumental problems and were re-observed by the XMM-{\it Newton}
between January and  February 2003. The EPIC (European Photon Imaging
Camera; Str\"uder et al. 2001; Turner et al. 2001) cameras were
operated in full frame mode with the thin filter applied.  

Both the F864 and SGP regions overlap with the
2dF Galaxy Redshift Survey
(2dFGRS\footnote{http://msowww.anu.edu.au/2dFGRS/}; Colless et
al. 2001) and the 2dF QSO Redshift Survey
(2QZ\footnote{http://www.2dfquasar.org}; Croom et al. 2001). Both the 
2dFGRS and 2QZ are large-scale spectroscopic campaigns that fully 
exploit the capabilities of the 2dF multi-fibre spectrograph on the
4\,m Anglo-Australian Telescope (AAT). These on-going
projects aim to obtain high quality spectra, redshifts and spectral
classifications for 250\,000 $bj<19.4$\,mag galaxies and 25\,000
optically selected $bj<20.85$\,mag QSOs. 

In addition to 2dF spectroscopy, the F864 region overlaps with 
the Sloan Digital Sky Survey (York et al. 2000). The  SDSS is an
on-going imaging and spectroscopic survey that aims to cover about $\rm
10\,000\,deg^2$ of the sky. Photometry is performed in 5 bands
($ugriz$;  Fukugita et al. 1996; Stoughton et al. 2002) to the
limiting magnitude $g \approx 23$\,mag, providing a uniform and
homogeneous multi-color photometric catalogue. The SDSS spectroscopic
observations will obtain spectra for over 1 million objects, including 
galaxies brighter than $r=17.7$\,mag, luminous red galaxies to
$z\approx0.45$ and colour selected QSOs (York et al. 2000; Stoughton
et al. 2002).  

Unlike the F864 area, the SGP region does not have complete and
homogeneous CCD photometric coverage. In the absence of good quality
photometry for this field we use photographic data from the APM survey
calibrated to the Johnson-Cousin $B$-band using CCD photometry
available for a subregion of the SGP field.  

\section{X-ray data}\label{sec_xray}

A full description of the X-ray data reduction and the generation of
the PN and MOS event files is presented by Georgakakis et al. (2003a).  To 
increase the signal--to--noise ratio and to reach fainter fluxes the
PN and the MOS event files have been combined into a single event list
using the {\sc merge} task of SAS.  Images have been extracted in the
spectral bands 0.5-8 (total), 0.5-2 (soft) and 2-8\,keV (hard) for
both the merged and the individual PN and MOS event files. We use the 
more sensitive (higher S/N ratio) merged image for source  extraction
and flux estimation, while the individual PN and MOS images are used
to calculate hardness ratios. This is because the interpretation of
hardness ratios is simplified if the extracted count rates  are from
one detector only.  

Source extraction is performed in the 0.5-8\,keV merged image using
the {\sc ewavelet} task of SAS with a detection threshold of
$5\sigma$. The extracted sources for each field were visually
inspected and spurious detections clearly associated with CCD gaps,
hot pixels  or lying close to the edge of the field of view were
removed. The final catalogue comprises a  total of 516 X-ray sources
to the limit $f_X(\rm 0.5 - 8 \,keV) \approx 10^{-14} \,erg \,s^{-1} 
\,cm^{-2}$. Source detection using the {\sc ewavelet} task of SAS with
a detection threshold of  $5\sigma$ is also performed in the
individual soft (0.5-2\,keV) and hard (2-8\,keV)  band merged images.
A total of 483 and 175 sources are detected in the individual 0.5-2
and  2-8\,keV  bands respectively to the $5\sigma$ detection
threshold. 

Count rates in the merged (PN+MOS) images as well as the
individual PN and MOS images are estimated within an 18\,arcsec
aperture. For the  background estimation we use the background maps
generated as a by-product of the {\sc ewavelet} task of  SAS. A small
fraction of sources lie close to masked regions (CCD gaps or hot
pixels) on either the MOS or the PN detectors. This may introduce 
errors in the estimated source  counts. To avoid this bias, the source
count rates (and hence the hardness ratios and the flux) are estimated 
using the detector (MOS or PN) with no masked pixels in the vicinity
of the source.   

To convert count rates to flux the Energy Conversion Factors
(ECF) of individual detectors are calculated assuming a power law
spectrum with $\Gamma=1.7$ and Galactic absorption $N_H=2\times
10^{20} \rm {cm^{-2}}$ appropriate for both the SGP and the F864
fields. The mean ECF for the mosaic of all three detectors is
estimated by weighting the ECFs of individual detectors by the
respective exposure time.  For the encircled energy correction,
accounting for the energy fraction outside the aperture within which 
source counts are accumulated, we adopt we adopt the calibration
 given by the {\it XMM-Newton} Calibration Documentation
\footnote{http://xmm.vilspa.esa.es/external/xmm\_sw\_cal/calib \\
/documentation.shtml\#XRT}.

\section{Optical photometric data}\label{sec_phot}
The F864 region overlaps with the SDSS (York et al. 2000) Early Data
Release (EDR; Stoughton et al. 2002;
http://www.sdss.org). Multiwaveband photometric observations
($ugriz$-filters; Fukugita et al. 1996) are available to the limiting 
magnitude $r\approx23$\,mag with the star-galaxy separation being
reliable to $r=21$\,mag. 
In the present study the colour transformations of Fukugita
et al. (1996) are used to convert SDSS $g$-band magnitudes to the
standard Johnson-Cousin $B$-band.    

In the  SGP region in the absence of homogeneous CCD photometry we 
use data from the APM survey. These are calibrated  to the standard
Johnson-Cousin $B$-band using CCD broad-band photometry available for
one of our X-ray pointings, SGP-2 [RA(J2000)=$00^{\rm h} 57^{\rm m}
00^{\rm s}$, Dec.(J2000)=$-27\degr 36\arcmin 00\arcsec$].    

Photometric observations of this field  in the Sloan $gri$
filters were carried out at the AAT in 2000 December 27 using the Wide 
Field Imager (WFI). The WFI was mounted at the prime focus of AAT
giving a pixel size of $\rm 0.225\,arcsec\,pixel^{-1}$ and a field of
view of $\rm 30\times30\,arcmin^{2}$. The total integration times were
2400\,s in all three bands split into four separate 600\,s
exposures. The observations were reduced following standard 
procedures using Starlink and IRAF tasks. Photometric calibration was
performed using standard stars from Landolt (1992). These provide the
instrumental zero point, the atmospheric extinction relation  and the
colour terms for the conversion from Sloan filters to the standard
Johnson-Cousin system used by Landolt. The uncertainty in the  zero
point estimated using these standard stars is $\pm0.05$\,mag in all
three bands. Source extraction and photometry is performed using the
SExtractor package (Bertin \& Arnouts 1996). Total Kron magnitudes are
measured in the $g$ and $r$ filters to estimate $g-r$ colours and to
transform $g$-band magnitudes into the standard Johnson-Cousin $B$-band
using the colour transformations derived above. A detailed description
of these observations including data reduction, source extraction and
catalogue generation will be presented in a forthcoming paper (Vallb\'e
et al. 2003, in preparation).     

Optical sources (both stars and galaxies) within the XMM/2dF survey
SGP region  are selected from the APM scans of the UKST $bj$
plates\footnote{http://www.ast.cam.ac.uk/$\sim$apmcat}. The APM
magnitudes are recalibrated to the Johnson-Cousin $B$-band using our
CCD photometry. 

Firstly, the APM and the CCD source catalogues are cross-correlated
using a 2\,arcsec matching radius to identify common objects. We find
843 overlapping sources of which 375 are galaxies and 468 are
classified stars by APM. These sources are used to estimate the
calibration curve, $B= f ( m_{bj})$, giving standard Johnson-Cousin
$B$-band magnitude as a function of uncalibrated APM magnitude
$m_{bj}$. Since the APM magnitudes are not corrected for emulsion
saturation the calibration curve is expected to be
non-linear. Following Maddox et al. (1990) we approximate $f(m_{bj})$
using 2nd and 3rd order polynomials for galaxies and stars
respectively (on the basis of the APM classification). 
The polynomial fit residuals for both galaxies and stars follow a
Gaussian distribution with  an rms of $\approx0.2$\,mag. This is 
the $1\sigma$ uncertainty of the calibrated APM magnitudes.

\section{Optical spectroscopic data}\label{sec_spec}

In addition to publicly available  spectroscopic data from 2dFGRS, 2QZ
and the SDSS spectroscopic surveys we have initiated our own follow-up
spectroscopic campaign of the XMM/2dF sources.   A detailed
description of these observations will be presented in a forthcoming
paper (Georgakakis et al. in preparation).     

In brief, X-ray sources in the F864 region of the XMM/2dF survey
with optical counterparts brighter than $B=22$\,mag were selected for
multi-fibre spectroscopy using 2dF at the prime focus of the AAT. The 2dF
consists of two spectrographs and two $1024 \times 1024$ thinned
Tektronix CCDs each receiving 200 fibres.  The fibres are
$\simeq$2\,arcsec in diameter resulting in 2 pixel wide spectra on the
detectors.

The data were obtained in service time mode during 2003 March 26.  
Due to poor weather conditions at the time of the observation the
total exposure   time was limited to 1\,h split into two half hour
integrations. The grating used was the 300B providing a dispersion of
$\rm 4.3\,\AA\,pixel^{-1}$ and a wavelength resolution of $\approx \rm
9\,\AA$  ($\simeq 2$\,pixels FWHM) over the range 3700--7900\AA. 

The data reduction was performed using the pipeline
reduction package  2DFDR developed for the
reduction of the 2dF data.  Fibre flat fields were employed to
determine the positions of the fibre spectra on each CCD frame 
and to flat-field the data. A CuArHe arc lamp exposure
was then used for the wavelength calibration. Redshifts were
determined  by visual inspection of the resulting spectra. Flux
calibration  has not been performed. This is due to the difficulty in
obtaining absolute flux calibration for the 2dF fibres which can
differ substantially in their throughput. This also applies to the
2dFGRS data. Therefore when presenting optical spectra (see Figure
\ref{fig_ngp_spec}) we plot raw counts as a function of wavelength.

\begin{figure*} 
\centerline{\psfig{figure=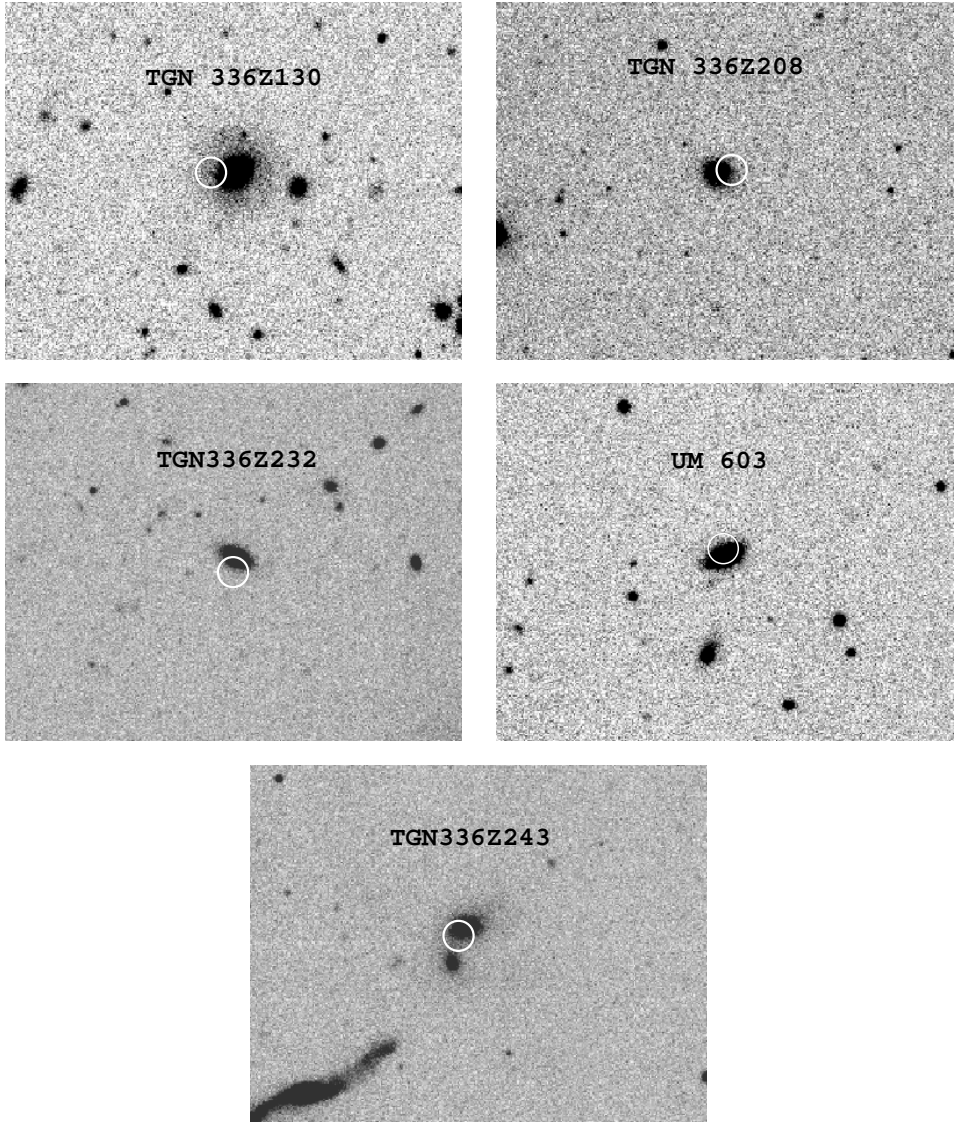,width=5in,angle=0}} 
\caption
 {Optical ($r$-band) images of the low X-ray--to--optical flux ratio
 sources identified with low redshift galaxies ($z<0.1$). These
 sources include the two  `normal' galaxy candidates (TGN\,336Z243,
 UM\,603) and the  three systems likely to host LLAGNs (TGN\,336Z130,
 TGN\,336Z208, TGN\,336Z232) of which one shows broad emission line
 optical spectrum. The optical spectra of these systems are shown in
 Figure \ref{fig_ngp_spec}. The position  of the
 X-ray centroid estimated by the {\sc ewavelet} task is overlaid on 
 the optical image. The white circles have radius of 4\,arcsec.
 }\label{fig_ngp}    
\end{figure*}

\section{Optical identification}\label{sec_id}
To optically identify the sources detected in the XMM/2dF survey we
follow  the method described by Downes et al. (1986) to calculate the
probability a given  candidate is the true identification using Bayes'
theorem: Consider an  optically detected candidate with magnitude $m$
at a distance $r$ from the X-ray position. Given the surface density
of objects brighter than $m$, $\Sigma(<m)$, the expected number of
candidates within $r$ is   

\begin{equation}
\mu=\pi\,r^{2}\,\Sigma(<m).
\end{equation}

\noindent Assuming that source positions are Poissonian, the probability
of at least one object brighter than $m$ within radius $r$ is

\begin{equation}
P=1-\exp(-\mu),
\end{equation}

\noindent which reduces to $\mu$ for $\mu<<$1. In this case, the
candidate is unlikely to be a chance association. In the present study
we apply an upper limit in the search radius, $r<7\rm\,arcsec$ and a
cutoff in the probability, $P<0.05$, to limit the optical
identifications to those candidates that are least likely to be
spurious alignments.  The background density of sources (both
galaxies and stars) in the Johnson-Cousin $B$-band is estimated using
the SDSS $g$-band converted to the $B$-band using the transformations
of Fukugita et al. (1996).  At magnitudes fainter than $B=22.5$\,mag 
where  SDSS is affected by incompleteness we use the $B$-band surface
density of Metcalfe et al. (1995). In the F864 region for the optical
identification we only consider SDSS sources brighter than
$B=22.5$\,mag to avoid the SDSS incompleteness at fainter
magnitudes. Similarly, in the SGP region we only consider APM sources
brighter than $B=21.5$\,mag. 

We propose 164 candidate optical identifications out of 291 0.5-8\,keV
selected sources in the F864 region and  57 identifications out of
223 X-ray  sources in the SGP region. 

The probability $P$ above is estimated under the assumption that the
source positions follow the Poisson distribution. To assess the
fraction of spurious optical identifications using the real spatial
distribution of sources we perform Monte Carlo simulations. Mock X-ray
catalogues are constructed by randomising the positions of the X-ray
sources within the area covered by the XMM-Newton observations. The
optical identification method is performed on the mock catalogues
using the same criteria ($P<0.05$, $\delta r<7$\,arcsec, $B<21.5$ or
$B<22.5$\,mag) as for the real sources. This procedure is
repeated 100 times. We find a spurious rate of $\approx6\%$ slightly
larger than the propability cutoff, $P$, assuming Poisson distribution
for the  source positions.

 \begin{figure*} 
 \centerline{\psfig{figure=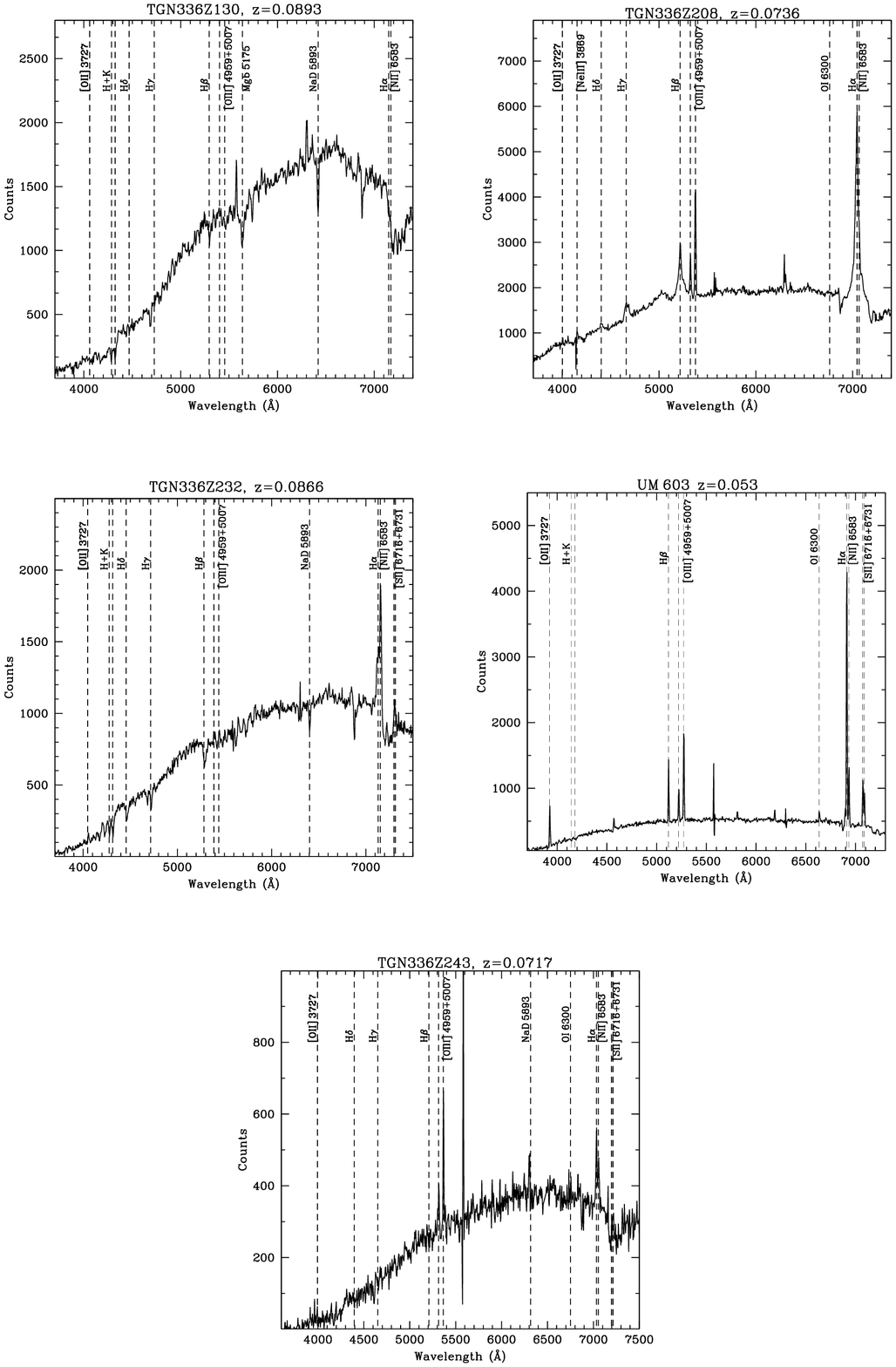,width=6in,angle=0}} 
 \caption
  {Optical spectra of the low X-ray--to--optical flux ratio  sources
  identified with low-$z$ galaxies ($z<0.1$). These sources include
  the two `normal' galaxy candidates (TGN\,336Z243, UM\,603)
  exhibitting narrow emission lines and  the  three systems 
  likely to host LLAGNs (TGN\,336Z130, TGN\,336Z208,
 TGN\,336Z232). The latter class comprises one galaxy 
  with broad emission line optical spectrum while the 
 remaining two show either absorption lines only or both absorption
 and weak narrow emission lines. 
  }\label{fig_ngp_spec}    
 \end{figure*}

\begin{table*} 
\footnotesize 
\begin{center} 
\begin{tabular}{llcc cccc cc}
\hline 
ID &
NAME &
$\alpha_X$ & 
$\delta_X$ & 
Offset &
$B$    &
Morphological  & 
$z$    &
Class  &
Comments$^{a}$ 
\\
 
 &
 &
(J2000)&
(J2000)&
(arcsec) &
(mag)    &
 type 
        &
         & 
         \\

1 &
 SDSS\,J134521.45-000121.04 &
 13  45   21.42  & --00   01    22.5 &
 1.6  &
 16.99 &
 6 &
 0.0$^1$ &
 star   &
 M-type\\

2 &
 TGN\,336Z130 & 
 13 45 15.31 & +00 15 15.9 &
 6.7 &
 17.63 & 
 3 &
 0.0893$^2$ & 
 E/S0  &
 A\\

3 &
 SDSS\,J134514.98-000047.91&
 13  45   14.98  & --00   00    47.6 &
 0.4  &
 13.69 &
 6 &
 --   &
 --  &
 --\\

4 &
 SDSS\,J134501.7-002401.74 &
 13  45    01.68  & --00  24     02.5 &
 0.9  &
 18.24 &
 6 &
 0.0$^1$    &
 star  &
 F/G-type\\

5 &
 SDSS\,J134433.57-000536.9 &
 13  44   33.81  & --00   05    37.0 &
 3.5  &
 19.97 &
 6 &
 0.0$^3$   &
 star  &
 M-type\\

6 &
 2QZ\,J134427.9-003029&
 13  44   27.98  & --00  30    32.3 &
 3.7  &
 19.44 &
 6 &
 1.374$^{1,4}$ &
 QSO  &
 BL \\

7 &
 SDSS\,J134425.94-000056.2 &
 13  44   25.97  & --00   00    55.3 &
 0.9  &
 18.95 &
 6 &
 1.097$^5$ &
 QSO  &
 BL \\

8 &
 TGN\,336Z208 & 
 13 43 51.13 & +00 04 38.8 &
 3.4  &
 17.62 &
 3 &
 0.0736$^2$ &
 Scd   &
 BL \\

9 &
 2QZ\,J134314.8+002528 &
 13  43   14.85  & +00  25    29.3 &
 0.4  &
 19.78  &
 6 &
 1.468$^4$ & 
 QSO &
 BL \\

10 &
 2QZ\,J134301.5-002951    &
 13  43    01.57  & --00  29    51.4 &   
 0.8  &
 20.17  &
 6 &
 2.062$^{1,4}$ & 
 QSO &
 BL \\

11 &
 SDSS\,J134235.91+002806.16  &
 13  42   35.91  & +00  28    10.0 &
 3.9 &
 13.81  &
 6 &
 0.0$^1$ & 
 star &
 F/G-type \\

12 &
 SDSS\,J134232.84-002008 &
 13  42   33.21  & --00  20     05.7 & 
 6.0  &
 17.48  &
 6 &
 -- & 
 -- &
 -- \\

13 &
 SDSS\,J134228.98+001947.5    &
 13  42   29.08  & +00  19    44.7 &
 3.1  &
 14.28  &
 6 &
 0.0$^1$ & 
 star &
 G-type \\

14 &
 TGN\,336Z232 &
 13 42 12.03 & --00 17 37.4  &
 2.8  &
 17.65 &
 3 &
 0.0866$^2$  &
 Sa &
 EA \\

15 &
 UM\,603 &
 13 41 37.85 & --00 25 55.3  &
 2.4 &
 16.86  &
 3 &
 0.053$^{1,6}$ &
 H\,II &
 NL\\

16 &
 2QZ\,J134133.6-00270 &
 13  41   33.82 & --00  27   02.8 &
 2.5  &
 19.84 &
 6 &
 1.341$^4$  &
 QSO &
 BL \\

17 &
 TGN\,336Z243 &
 13 41 33.22 & --00 24 34.2  &
 3.1  &
 17.63 &
 3 &
 0.0717$^2$  &
 Sa &
 NL \\

18 &
 SDSS\,J134125.7-002208.06 &
 13  41   25.69 & --00  22     07.8 &
 0.4  &
 20.05 &
 6 &
 0.0$^1$  &
 star &
 M-type \\

19 &
 SDSS\,J134056.52+003156.29 &
 13  40   56.52 & +00  31    57.9 &
 1.6  &
 14.02 &
 6 &
 0.0$^1$  &
 star &
 K-type \\

20 &
 XMM2DF\,J005929.63-275316.07 &
 00  59   29.63 & --27  53    16.1  &
 2.6  &
 17.1 &
 6 &
 --  &
 --  &
 -- \\

21 &
 B005615.26-275548.8 &
 00  58   40.48 & --27  39    41.6 &
 5.3  &
 20.1 &
 3 &
 --  &
 --  &
 -- \\

22 &
 XMM2DF\,J005822.88-274014.03 &
 00  58   22.88 & --27  40    14.0 &
 1.1  &
 15.2 &
 6 &
 0.0$^3$  &
 star &
 F/G-type \\

23 &
 XMM2DF\,J005803.14-280856.05 &
 00  58    03.14 & --28   08    56.1 &
 1.5  &
 15.1 &
 6 &
 --  &
 -- &
 -- \\

24 &
 2QZ\,J005734.9-272828 &
 00  57   34.92 & --27  28    29.1 &
 1.3  &
 19.0 &
 6 &
 2.189$^4$  &
 QSO &
 BL \\

25 &
 2QZ\,J005701.1-272800 &
 00  57    01.04 & --27  28   01.6 &
 2.2  &
 19.7 &
 6 &
 0.825$^4$  &
 QSO &
 BL \\

26 &
 XMM2DF\,J005637.45-272717.47  &
 00  56   37.45 & --27  27    17.5 &
 1.7  &
 15.3 &
 6 &
 --  &
 --  &
 -- \\

\hline
\multicolumn{10}{l}{$^a$A: absorption lines; NL: Narrow emission lines;
BL: Broad emission lines; EA: absorption+emission lines} \\
\multicolumn{10}{l}{1: XMM/2dF spectroscopic program; 2: 2dFGRS; 3:
Griffiths et al. (1995); 4: 2QZ; 5: SDSS; 6: Terlevich et al. (1991)}\\
\end{tabular} 
\end{center} 
\caption{
Low X-ray--to--optical flux ratio sources in the XMM/2dF survey.  
}\label{tbl2} 
\normalsize  
\end{table*}

\begin{table*} 
\footnotesize 
\begin{center} 
\begin{tabular}{lcc c ccc} 
\hline 
ID &
\multicolumn{2}{c}{count rate$^{a}$} &
$f_X(\rm 0.5-8\,keV)$$^{b}$ &
$\log f_X/f_{opt}$ &
$L_X(\rm 0.5-8\,keV)$ $^{b}$&
 HR$^{a}$ 
 \\

  &
\multicolumn{2}{c}{($\rm \times 10^{-3}\,cnts\,s^{-1}$)} &
($\times 10^{-14}\,\rm erg\,s^{-1}\,cm^{-2}$) &
  &
($\rm\,erg\,s^{-1}$) &
\\
    
  &
  0.5-2\,keV &
  2-8\,keV &
      &
      &
      &
      \\  

1 & $43.90\pm8.84$ &  $<13.8$ & $12.4\pm1.74$ & $-1.22$ &  
 -- &  $<-0.52$ 
\\

2$^{c,\star}$ & $3.13\pm0.93$ &  $<3.50$ &  $3.68\pm1.02$ & $-1.50$ & 
$(8.38\pm2.32)\times10^{41}$ &  $<0.05$ 
\\

3 & $160.00\pm12.80$ &  $7.97\pm3.86$ &  $48.49\pm2.65$ & $-1.95$ &
-- & $-0.91\pm0.11$ 
\\

4 & $6.21\pm2.03$ &  $5.53\pm1.94$ &  $3.91\pm0.56$ & $-1.22$ &
-- & $-0.06\pm0.24$ 
\\ 

5 & $2.13\pm2.14$ &  $<8.69$ &  $1.03\pm0.77$ &  $-1.11$ &
-- &  $<0.61$ 
\\ 

6 & $<7.77$ &  $<6.57$ &  $0.82\pm0.41$ &  $-1.42$ &
$(8.53\pm4.26)\times10^{43}$ &
-- 
\\

7 & $11.00\pm2.95$ &  $<10.30$ &  $3.05\pm0.57$ & $-1.05$ &
$(1.89\pm0.36)\times10^{44}$ &  $<-0.03$ 
\\


8$^{c,\star}$ & $10.30\pm1.59$ &  $3.34\pm0.98$ &  $12.00\pm1.61$ &$-0.99$ &
$(1.81\pm0.24)\times10^{42}$ &  $-0.51\pm0.15$ 
\\

9$^c$ & $1.27\pm0.51$ &  $<1.37$ &  $1.29\pm0.51$ & $-1.09$ &
$(1.55\pm0.62)\times10^{44}$ &  $<0.04$ 
\\

10 & $3.55\pm2.74$ &  $<13.3$ &  $0.91\pm0.73$ & $-1.09$ &
$(2.38\pm1.91)\times10^{44}$ &  $<0.58$ 
\\

11 & $4.98\pm2.37$ &  $<6.05$ &  $2.39\pm0.62$ & $-3.21$ &
-- &  $<0.10$ 
\\

12 & $0.84\pm3.17$ &  $<22.00$ &  $3.11\pm0.88$ &  $-1.62$ &
-- &  $<0.93$ 
\\

13 & $13.60\pm3.86$ &  $<7.44$ &  $3.88\pm0.85$ & $-2.81$ &
-- &  $<-0.29$ 
\\

14$^{\star}$ & $6.06\pm3.70$ &  $<15.69$ &  $3.64\pm1.08$ & $-1.49$ &
$(7.82\pm2.32)\times10^{41}$ &  $<0.44$ 
\\

15 & $8.47\pm2.46$ &  $<6.32$ &  $1.95\pm0.52$ & $-2.08$ &
$(1.52\pm0.40)\times10^{41}$ &  $<-0.15$ 
\\

16 & $2.93\pm1.84$ &  $1.56\pm1.50$ &  $1.52\pm0.48$ & $-0.99$ &
$(1.49\pm0.48)\times10^{44}$ &  $-0.31\pm0.55$ 
\\

17 & $<9.76$ &  $<5.42$ &  $1.31\pm0.43$ & $-1.95$ &
$(1.89\pm0.63)\times10^{41}$ &  -- 
\\

18 & $3.96\pm1.78$ &  $<3.77$ &  $0.84\pm0.39$ & $-1.17$ &
-- &  $<-0.03$ 
\\

19 & $132.00\pm8.25$ &  $3.51\pm1.93$ &  $37.20\pm1.64$ & $-1.94$ &
-- &
$-0.95\pm0.09$ 
\\

20 & $7.49\pm2.18$ &  $<5.26$ &  $1.99\pm0.50$ &  $-1.99$ &
-- &  $<-0.18$ 
\\

21 & $1.66\pm1.50$ &  $<3.89$ &  $0.34\pm0.35$ & $-1.55$&
-- & $<0.40$ 
\\

22 & $68.70\pm7.06$ &  $-0.96\pm1.29$ &  $22.50\pm1.47$ &$-1.69$ &
 -- & $-1.03\pm0.15$ 
\\

23 & $105.00\pm8.30$ &  $16.40\pm3.82$ &  $42.30\pm2.24$ & $-1.43$ &
-- &
$-0.73\pm0.09$ 
\\

24 & $6.70\pm2.61$ &  $<7.80$ &  $2.87\pm0.65$ & $-1.06$ &
$(8.53\pm1.94)\times10^{44}$ &  $<0.08$ 
\\

25 & $<6.35$ &  $4.08\pm2.06$ &  $1.88\pm0.51$ & $-0.98$ &
$(6.01\pm1.66)\times10^{43}$ &  $>-0.22$ 
\\

26 & $24.00\pm4.20$ &  $6.55\pm2.68$ &  $10.20\pm1.05$ & $-2.00$ &
-- & $-0.57\pm0.18$ 
\\

\hline
\multicolumn{6}{l}{$^a$Count rates and hardness ratios are from EPIC-PN
unless expliciltly stated otherwise.} \\
\multicolumn{6}{l}{We define $\rm HR=(H-S)/(H+S)$ as
in equation \ref{eq2}.} \\
\multicolumn{6}{l}{$^b$Fluxes and luminosity estimates are from the
merged images  unless expliciltly stated otherwise} \\
\multicolumn{6}{l}{$^c$Count rates, hardness ratios and X-ray fluxes
are from MOS} \\
\multicolumn{6}{l}{$^\star$Classified LLAGN: $L_X \la 10^{42}\rm\,
erg\,s^{-1}$ and optical/X-ray properties suggesting AGN activity} \\

\end{tabular} 
\end{center} 
\caption{X-ray properties of low X-ray--to--optical flux ratio sources
in the XMM/2dF survey}\label{tbl3}  
\normalsize  
\end{table*} 

\section{The sample}\label{sample}
The sample used in the present study is compiled from the $5\sigma$
threshold 0.5-8\,keV source catalogue by identifying and selecting low
X-ray--to--optical flux ratio sources, $\log f_X/f_{opt}<-0.9$. The
X-ray--to--optical flux ratio is estimated from the 0.5-8\,keV flux
$f(0.5-8\,{\rm keV})$ and the $B$-band magnitude according to 
the relation 
\begin{equation}\label{eq1}
\log\frac{f_X}{f_{opt}} = \log f(0.5-8\,{\rm keV}) +
0.4\,B +4.89.
\end{equation}
The equation above is derived from the X-ray--to--optical flux
ratio definition of Stocke et al. (1991) that involved 0.3-3.5\,keV
flux and $V$-band magnitude. These quantities are converted to
0.5-8\,keV flux and $B$-band magnitude assuming a mean colour
$B-V=0.8$ and a power-law X-ray spectral energy distribution with
index  $\Gamma=1.7$. 

The 0.5-8\,keV selected sample with $\log f_X/f_{opt}<-0.9$ is
presented in Table \ref{tbl2} which has the following format:

{\bf 1.} Name of the most probable optical counterpart of the
X-ray source. If a name is not available we use the prefix ``XMM2DF''
followed by the RA and DEC coordinates of the X-ray source in J2000.    

{\bf 2-3.} Right ascension ($\alpha_X$) and declination ($\delta_X$) of
the X-ray source position in J2000.

{\bf 4.} Offset in arcseconds between the X-ray source centroid estimated 
by the {\sc ewavelet} task of {\sc sas} and the optical source centre. We 
note that the X-ray source centroid does not always coincide with 
the peak of the X-ray emission. 

{\bf 5.} Optical $B$-band magnitude.

{\bf 6.} Optical morphology of the source: 3 is for extended optical light
profile (i.e. galaxy) and 6 is for unresolved optical
light profile (i.e. star or QSO). The optical morphology
classification is from the SDSS and the APM for sources in the F864
and SGP regions respectively. 

{\bf 7.} Redshift. The source from which the redshift estimate was
obtained is also listed in Table \ref{tbl2}. 

{\bf 8.} Spectral classification of the optical counterpart. Spectral
classifications are obtained from the same source as the redshifts
listed in the previous column. 

{\bf 9.} Comments on the observed optical spectral features: {\bf A:}
absorption lines; {\bf NL:} narrow emission lines; {\bf BL:} broad
emission lines; {\bf EA:} both absorption and emission lines. If the
most probable counterpart is a star we give its  spectral type. 

Table \ref{tbl3} presents the X-ray properties of the low $\log f_X/
f_{opt}$ sample. We list:

{\bf 1.}  0.5-2 and 2-8\,keV count rates. Sources that are not detected in
the 0.5-2 or 2-8\,keV spectral bands above the $5\sigma$ threshold are
assigned an upper limit  ($3\sigma$) assuming Poisson statistics. A small
number of sources are only detected in the 0.5-8\,keV band but 
not in the individual 0.5-2 and 2-8\,keV bands.  For these sources an
upper limit is estimated  to both their 0.5-2 and 2-8\,keV count rates. 
 
{\bf 2.}  0.5-8\,keV X-ray flux in $\rm erg\,s^{-1}\,cm^{-2}$. 

{\bf 3.} 0.5-8\,keV X-ray luminosity in $\rm erg\,s^{-1}$, if a redshift
is available.

{\bf 4.} Hardness ratio, $\rm HR$, defined 
\begin{equation}\label{eq2}
\rm HR = \frac{\rm RATE(2080)-RATE(0520)}{\rm RATE(2080)+RATE(0520)},
\end{equation}
where $\rm RATE(0520)$ and $\rm RATE(2080)$ are the count rates in
the 0.5-2 and 2-8\,keV spectral bands respectively. Upper and lower
limits are  for sources that are not detected in the hard and soft
bands respectively to the {\sc ewavelet} detection threshold of 
$5\sigma$. In the case of soft/hard band non-detection a $3\sigma$
upper limit is estimated assuming Poisson statistics.  

We note that for all the X-ray sources presented here the probability
of an individual optical counterpart being spurious coincidence is small 
$P\la6\times10^{-3}$. Figure \ref{fig_ngp} shows the optical
images of the low-$z$ sources ($z<0.1$) with the position of the X-ray
centroid (estimated by the {\sc ewavelet} task) overlaid. The optical
spectra of the same sources are presented in Figure
\ref{fig_ngp_spec}.  

\begin{figure} 
\centerline{\psfig{figure=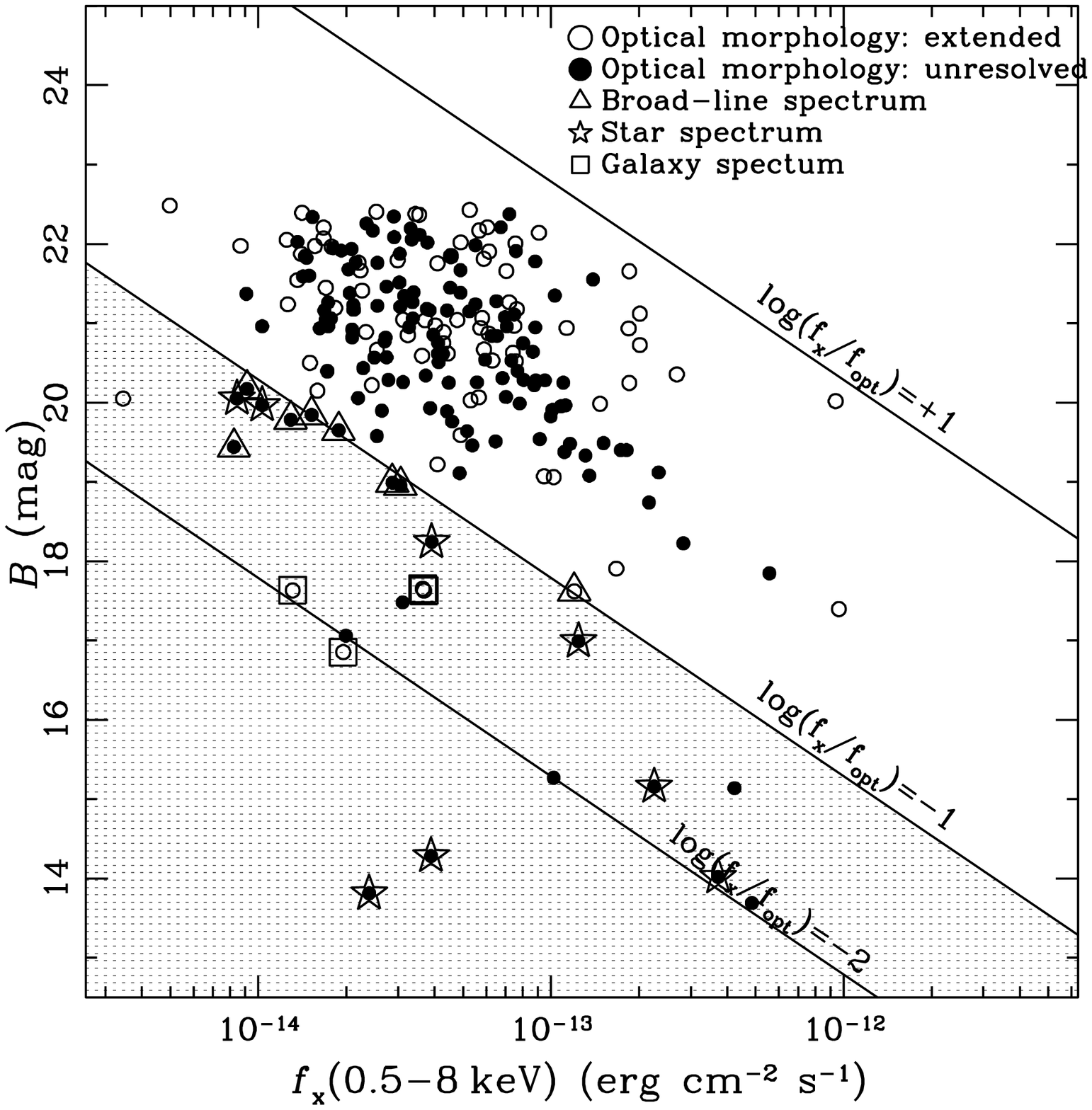,width=3.5in,angle=0}} 
\caption
 {$B$-band magnitude against 0.5-8\,keV flux. The $\log
 f_X/f_{opt}<-1$ region studied here is shaded. Open circles are
 0.5-8\,keV X-ray detections with extended optical light
 profile. Filled circles are 0.5-8\,keV X-ray detections with stellar-like
 optical light profile. A triangle on top of a symbol indicates broad
 line optical spectrum. A square on top of a symbol is for sources
 with optical spectra dominated by light from the host galaxy
 (e.g. narrow-emission and/or absorption lines). A star
 arround a symbol indicates sources with Galactic star optical
 spectra. The lines indicate constant X-ray--to--optical
 flux ratios of +1, --1 and --2. The lines $\log f_X/f_{opt}=\pm1$
 delineate the region of the parameter space occupied by powerful
 AGNs.  The 
 X-ray--to--optical flux ratio is defined in section \ref{sample}.
 }\label{fig_fxfo}    
\end{figure}

\begin{figure} 
\centerline{\psfig{figure=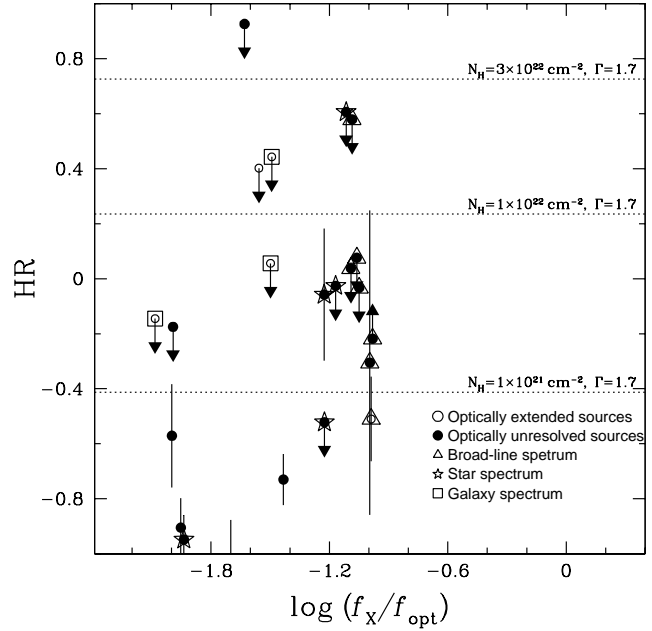,width=3.5in,angle=0}} 
\caption
 {Hardness ratio against X-ray--to--optical flux ratio as defined in
 section \ref{sample} (eq. \ref{eq1}).  The symbols are the same as in
 Figure \ref{fig_fxfo}.  
 Upper and lower
 limits are for sources that are not detected in the hard and soft
 bands respectively to the {\sc ewavelet} detection threshold of
 $5\sigma$. In the case of soft/hard band non-detection a $3\sigma$
 upper limit to the count rate is estimated assuming Poisson
 statistics.     
 }\label{fig_hr}    
\end{figure}

Figure \ref{fig_fxfo} plots $B$-band magnitude against 0.5-8\,keV
X-ray flux for both low $f_X/f_{opt}$ sources and the whole X-ray
selected sample. The  $\log (f_X/f_{opt})=\pm1$ lines in this figure
delineate the region of the parameter space occupied by AGNs. Figure 
\ref{fig_hr} plots the hardness ratio against X-ray--to--optical flux
ratio for the low X-ray--to--optical flux ratio sources.

Notes on individual selected sources are presented in Appendix
\ref{app1}. The classification of the present sample into different
classes is performed  on the basis of their optical spectroscopic
(i.e. spectral features), photometric (i.e. resolved or point-like
sources) and X-ray properties (i.e. X-ray luminosity,
X-ray--to--optical flux ratio, HR). More information about the
source classification can be found in  Appendix
\ref{app1}. The present sample of low X-ray--to--optical flux ratio
sources to the  limit $f(\rm 0.5 - 8 \,keV) \approx 10^{-14}
\,erg\,s^{-1}\,cm^{-2}$  comprises: {(i)} 8 spectroscopically
confirmed Galactic stars, {(ii)} 7 broad-line  QSO, {(iii)} 3 LLAGNs
of which  one exhibits broad lines (TGN\,336Z208) while the other two
show  absorption and/or narrow emission lines (TGN\,336Z130,
TGN\,336Z232),  {(iv)} 2 `normal' galaxy candidates (UM\,603,
TGN\,336Z243) and {(v)} 6 unclassified sources with no optical
spectroscopic  information. Most of the objects in the latter class
are likely to be Galactic stars on the basis of their  optical and
X-ray properties.

We note that by LLAGN (e.g. class (iii) above) we refer to sources
that show evidence for AGN activity and have X-ray luminosity $L_X\la
10^{42}\rm erg\,s^{-1}$ (e.g. Elvis,  Soltan \& Keel 1984). Indeed, 
the X-ray sources classified LLAGNs in our sample are nearby galaxies
($z\la0.1$) with X-ray luminosities $L(\rm 0.5-8\,keV) \approx 10^{42}
\,erg\,s^{-1}$ lower than those of distant QSOs. Two out of three
(TGN\,336Z130, TGN\,336Z232) have $\log f_X /f_{opt}\approx-1.5$ and
optical spectra exhibiting both absorption and narrow emission lines.

The candidate `normal' galaxies in the present sample have  $L(\rm
0.5-8\,keV) \approx 10^{41} \,erg\,s^{-1}$, $\log f_X
/f_{opt}\approx-2$ and narrow emission line optical spectra. These
values are elevated  compared to quiescent spirals  
($L(\rm 0.5-8\,keV) \approx 10^{40} \,erg\,s^{-1}$, $\log f_X
/f_{opt}\approx-3$; Hornschemeier et al. 2002; Georgakakis 
et al. 2003a; Hornschemeier et al. 2003)  and typical to those of
actively star-forming galaxies (e.g. Moran, Lehnert \& Helfand
1999; Alexander et al. 2002; Bauer et al. 2002; Georgakakis 
et al. 2003b).  We note that although the X-ray and optical properties
of `normal' galaxy candidates are consistent with stellar origin for
the X-ray emission  we cannot exclude the possibility of LLAGN.  

\begin{figure} 
\centerline{\psfig{figure=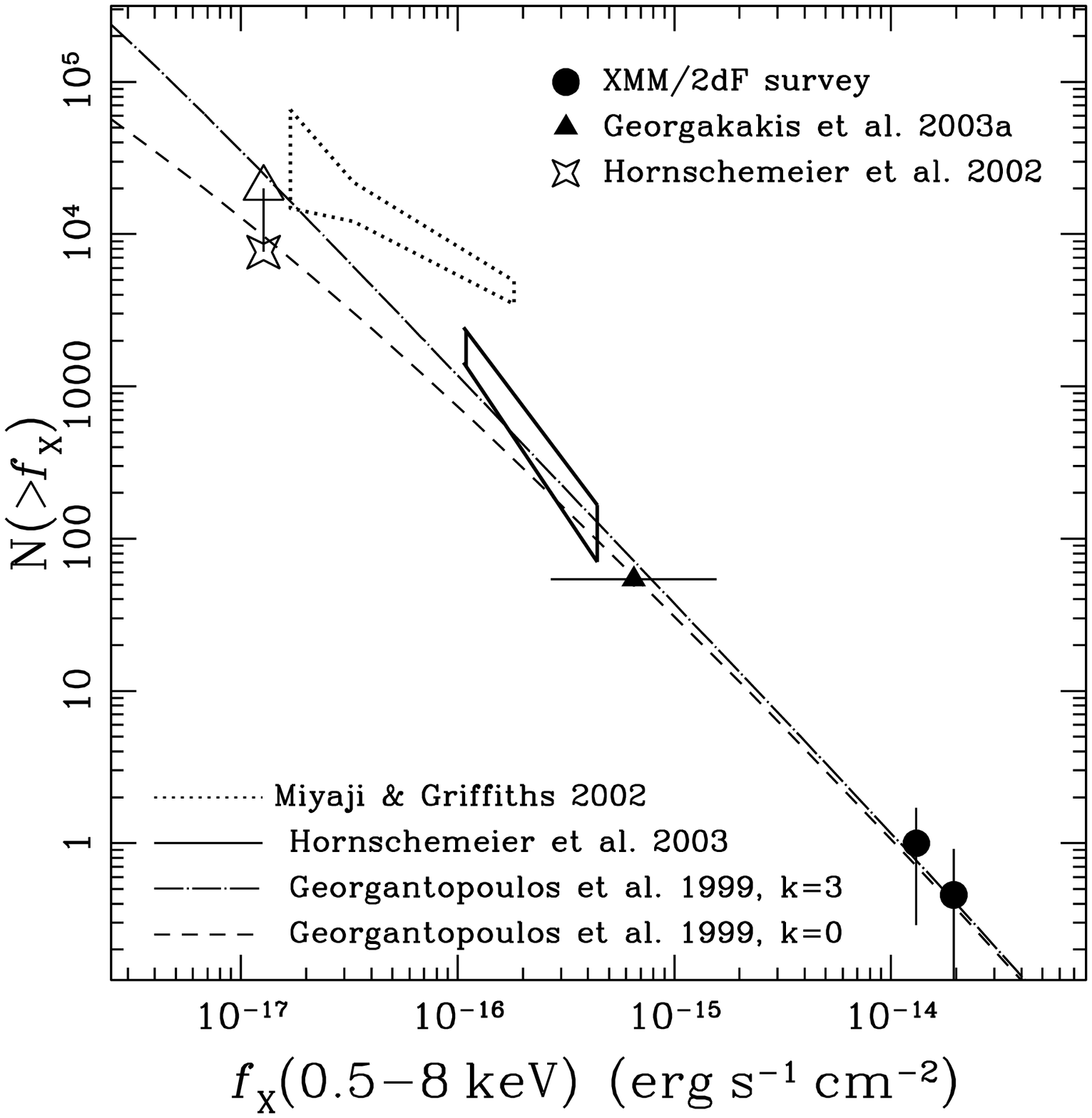,width=3.5in,angle=0}} 
\caption
 {Cumulative `normal' galaxy counts. Filled circles are the
 `normal' galaxy candidates in the present study. The triangle
 represents the constraints from the stacking analysis results of
 Georgakakis et al. (2003a). The solid lined rectangle marks the 
 region occupied by the source counts of Hornschemeier et
 al. (2003). The star is an upper limit to $\log N - \log S$ from the
 stacking analysis results of Hornschemeier et
 al. (2002). 
 The model predictions of the Georgantopoulos
 et al. (1999) X-ray luminosity function of H\,II galaxies assuming no
 evolution (dashed line) and luminosity evolution of the form
 $L_X\sim(1+z)^{3}$ respectively (dot-dashed line) are also
 plotted. The dotted line is the fluctuation analysis constraints of
 Mijayi \& Giffiths (2002) and should be regarded as an upper limit to
 the   `normal' galaxy counts.   
 }\label{fig_logn}    
\end{figure}

\section{Discussion}\label{discussion}
In the present study we use the wide area XMM/2dF survey to study  
low X-ray--to--optical flux ratio sources ($\log f_X / f_{opt} < -0.9$)
to the survey limit $f(\rm
0.5-8\,keV)\approx10^{-14}\,erg\,s^{-1}\,cm^{-2}$. This  
translates  to  a soft band  flux of $f(\rm 0.5-2\,keV)\approx
5\times 10^{-15}\,erg\,s^{-1}\,cm^{-2}$ (assuming $\Gamma=1.7$)
comparable to previous {\it ROSAT} PSPC surveys.  Out of 26 sources
with $\log f_X /  f_{opt} < -0.9$ we find:  (i) 8 spectroscopically
confirmed Galactic stars, (ii) 7 broad-line  QSO, (iii) 3 LLAGNs (for
a definition of LLAGN class see section \ref{sample})
of which one shows broad optical lines and the other two 
have absorption and/or narrow emission lines, (iv) 2 `normal' galaxy
candidates and (v) 6 unclassified sources with
no  optical spectroscopic information most of which are likely to be 
Galactic stars. The dominant populations to the flux limit of the
present survey are Galactic stars and broad line QSOs with only a
small fraction  of LLAGNs and `normal' galaxy candidates. The
relatively mix of the low X-ray--to--optical flux ratio population
found here is in fair agreement with the {\it ROSAT} HRI  results of
Lehmann et al. (2001) with the exception of the larger fraction of
galaxy groups identified by these authors. However, all the  galaxy
groups in the Lehmann et al. (2001) sample have fluxes $f_X(\rm
0.5-2\,keV) < 5\times10^{-15} \,erg\,s^{-1}\,cm^{-2}$, i.e. below the
sensitivity limit of the present survey. 

An interesting result from the present study is the identification of
two `normal' galaxy candidates. 
Recent deep {\it Chandra} surveys demonstrated
beyond any doubt the appearance of `normal' galaxies at faint X-ray
fluxes ($\rm \la 10^{-16}\,erg \,s^{-1} \,cm^{-2}$; Hornschemeier et
al. 2002; Bauer et al. 2002). Hornschemeier et al. (2003) used the 2\,Ms
CDF-N survey to  compile the first large X-ray selected sample of
distant (median redshift $\approx0.3$) quiescent galaxies with $\log
f_X/f_{opt}\la -2.3$. They  demonstrated that these systems comprise a
non-negligible fraction ($\approx15\%$) of the X-ray population at
$f(\rm 0.5-2\,keV) \approx 2\times10^{-17}\,erg \,s^{-1} \,cm^{-2}$
and that their  $\log N - \log S$ rises much steeper than the general
X-ray source population.  The majority of the spectroscopically
identified sources in the Hornschemeier et al. (2003) sample are
emission-line galaxies most likely to be star-forming spirals in the
redshift range $0.06 \la z \la 0.85$. 

Our shallow wide area XMM/2dF survey complements the Hornschemeier et
al. (2003) study by constraining the `normal' galaxy $\log N - \log S$
at much brighter fluxes $f(\rm 0.5-8\,keV) \approx 10^{-14}\,erg
\,s^{-1} \,cm^{-2}$, albeit with poorer statistics. Figure
\ref{fig_logn}  plots the cumulative `normal' galaxy
counts from the present sample (i.e. the two `normal'   galaxies
identified here) and the CDF-N survey (Hornschemeier et
al. 2003). Clearly, our surface density constraints suffer from small
number statistics (e.g. only  two `normal'   galaxy candidates). We
note however, that the XMM/2dF survey with its wide field of view
provides the only constaints todate to the surface density of the
rare X-ray selected `normal'  galaxies at the flux limit of $\approx
10^{-14}\rm\, erg\, s^{-1}\, cm^{-2}$.

Extrapolating the Hornschemeier et al. (2003) results to brighter flux
limits using the best fit power-law they adopt to describe their $\log
N - \log S$ we find that our source counts are higher by a factor of
$\approx5$. However, taking into account the uncertainties 
in both our sample and that of Hornschemeier et  al. (2003) there is
agreement between the two studies. Also, we note that
Hornschemeier et al. (2003) select only sources with  $\log
f_X/f_{opt} \la -2.3$, i.e. quiescent low star-formation systems. This
conservative selection threshold was chosen to avoid LLAGN
contamination but is also likely to miss actively star-forming
galaxies expected to have $-1 \la \log f_X/f_{opt} \la -2$ (Alexander
et al. 2002; Bauer et al. 2002).

Also shown in Figure \ref{fig_logn}   
are the constraints from the stacking analysis results of Georgakakis
et al. (2003a). These authors used the XMM/2dF survey to estimate the
mean X-ray properties of optically selected 2dFGRS spirals/ellipticals
at a mean redshift of $z\approx0.1$ by applying stacking methods.  
A statistically significant stacking signal was found for both the
elliptical and the spiral galaxy sub-samples providing an estimate of
the mean X-ray flux of these systems. We plot the surface density of
the spiral galaxy sample used by Georgakakis et al. (2003a) at the
mean X-ray flux of these systems estimated via stacking. This 
point is lower limit since it represents only a subset of the `normal'
galaxy population at the given flux limit.   

The fluctuation analysis results of Miyaji \& Griffiths (2002) are
shown in Figure \ref{fig_logn}. The $\log N - \log S$ constraints of
these authors refer to the whole X-ray population (e.g. including AGNs)
and not just the `normal' galaxy sub-sample. Nevertheless, it has been
suggested that `normal' galaxies are likely to outnumber AGNs at faint
fluxes  (Hornschemeir et al. 2003) and therefore the Miyaji \&
Griffiths (2002) results are likely to be relevant to `normal'
galaxies at faint fluxes. In any  case, the constraints provided by
these authors  are an upper limit to the `normal' galaxy counts.  

The model  $\log N - \log S$ prediction using the X-ray luminosity
function derived by Georgantopoulos, Basilakos \& Plionis (1999) for
H\,II galaxies is also plotted in Figure
\ref{fig_logn}. These authors convolved the local optical luminosity
function of the Ho, Filippenko \& Sargent (1995) sample with the
corresponding $L_X-L_B$ relation based on {\it Einstein} data to
derive the X-ray luminosity function of different galaxy types. The
galaxy classification scheme of the Ho  et al. (1995) sample is 
highly reliable since it is based on high S/N nuclear optical spectra
(Ho et al. 1997). Different X-ray luminosity evolution
scenarios of the form $L_X\sim(1+z)^{k}$ are
plotted. The difference between these two models at the flux range
plotted here is small because the predicted mean redshift of
star-forming galaxies at the flux limit $f(\rm 0.5 - 8 \,keV)
\approx 10^{-16}\,erg \,s^{-1} \,cm^{-2}$ remains low. This is in
agreement with the mean redshift of $\approx0.3$ of the
Hornschemeier et  al. (2003) deep `normal' galaxy sample.

The $k=3$ evolutionary model is in fair agreement with  
both the Hornschemeier et  al. (2003) and our $\log N - \log S$
estimates. Although the  Georgantopoulos et al. (1999)
model is flatter (slope$=-1.5$) than  the Hornschemeier et
al. (2003)  $\log N - \log S$ (slope=$-1.74^{+0.28}_{-0.30}$), within 
the $1\sigma$ uncertainties there is fair agreement.
Also, the $k=3$ model is in better agreement with the
Hornschemeier et  al. (2003) observations over a wider flux range
compared to  the $k=0$ 
model. Although the difference between the two models is marginal in
the flux range covered by the Hornschemeier et  al. (2003) data the
evidence above   provides weak evidence for X-ray evolution of
star-forming systems. As already discussed the mean redshift of
`normal'  galaxies at the flux limits probed by Hornschemeier et
al. (2003) is low and therefore any evolutionary effects are expected
to be small. Deeper X-ray observations are required to detect `normal' 
galaxies at moderate and high redshifts to better constrain their
evolution.

\section{Conclusions}\label{conclusions}
In this paper we employ a wide field ($\rm 2.5\,deg^{2}$) shallow
($f_X\rm (0.5 - 8\,keV)\approx 10^{-14} \, erg \, s^{-1}$) XMM-{\it
Newton} survey to investigate the nature of low X-ray--to--optical
flux ratio sources, $\log f_X / f_{opt}<-0.9$. Our sample comprises 26
objects of which 20 have optical spectroscopic information. The
majority of the sources without spectroscopic information have
X-ray/optical properties that strongly suggest Galactic stars. Most of
the spectroscopically identified systems are Galactic  stars (total of  
8) and broad line AGNs (total of 8).  

We also find four sources with  narrow emission and/or absorption
line optical spectra. Two of them have X-ray/optical properties
suggesting LLAGN activity. 
The remaining two sources have narrow emission
line optical spectra, X-ray luminosities $L_X\rm \approx
10^{41}\,erg\,s^{-1}$ and X-ray--to--optical flux ratios $\approx-2$
suggesting `normal' galaxies powered by star-formation activity.   

The small number of `normal' galaxies found in our wide field  XMM/2dF
survey is in agreement with the results from the 1\,Ms ROSAT HRI 
survey of the Lockman Hole. Despite  the poor
statistics the estimated number density of `normal' galaxies at the
flux limit probed here is consistent with  the $\log N / \log S$ of
deeper {\it Chandra} surveys  extrapolated to bright fluxes. Using the
X-ray luminosity function of local star-forming galaxies we find that
the predicted $\log N - \log S$ is in fair agreement with both our
shallow and the deeper {\it Chandra} samples. The pure luminosity
evolution model is in better agreement with the observations compared
to the no-evolution prediction providing some evidence for X-ray
evolution of star-forming spirals. However, the difference between the
two models even at the flux limits of the 2\,Ms Chandra Deep Field 
North survey is small and does not allow firm conclusions to be drawn.
This is because of the low mean redshift ($z\approx0.3$) of the
`normal' galaxies probed by this ultra deep survey.

\section{Acknowledgments}
 We thank the anonymous referee for valuable comments and
 suggestions. This work is jointly funded by the European Union  
 and the Greek Government  in the framework of the programme
 ``Promotion of Excellence in Technological Development and Research'',
 project ``X-ray Astrophysics with ESA's mission XMM''.  

 We acknowledge use of the 100k data release of the 2dF Galaxy   
 Redshift Survey. The 2dF QSO Redshift Survey (2QZ) was compiled by
 the 2QZ survey team from observations made with the 2-degree Field on
 the Anglo-Australian Telescope.  

 Funding for the creation and distribution of the SDSS Archive has
 been provided by the Alfred P. Sloan Foundation, the Participating
 Institutions, the National Aeronautics and Space Administration, the
 National Science Foundation, the U.S. Department of Energy, the
 Japanese Monbukagakusho, and the Max Planck Society. The SDSS Web
 site is http://www.sdss.org/. The SDSS is managed by the
 Astrophysical Research Consortium (ARC) for the Participating
 Institutions. The Participating Institutions are The University of
 Chicago, Fermilab, the Institute for Advanced Study, the Japan
 Participation Group, The Johns Hopkins University, Los Alamos
 National Laboratory, the Max-Planck-Institute for Astronomy (MPIA),
 the Max-Planck-Institute for Astrophysics (MPA), New Mexico State
 University, University of Pittsburgh, Princeton University, the
 United States Naval Observatory, and the University of Washington. 

\appendix 
\section{Notes on selected sources}\label{app1}

{\bf TGN\,336Z130:} The X-ray centroid lies $\rm \approx7\,arcsec$
away from the position of the optical galaxy. The probability
of chance coincidence is small
$P=0.6\%$. Also, inspection of the X-ray centroid overlaid on
the SDSS optical image in Figure  \ref{fig_ngp} provides additional
evidence that  TGN\,336Z130 is likely to be the correct counterpart of
the X-ray source. In what follows we assume that TGN\,336Z130 is the
correct identification. The 2dFGRS optical spectrum of this
$B=17.6$\,mag systems exhibits absorption lines only suggesting an
early type galaxy, most likely an E/S0. The X-ray--to--optical flux
ratio of  $\approx-1.5$ and the X-ray luminosity $L_X (\rm 0.5-8\,keV)
\approx 8\times 10^{41} \rm \, erg\, s^{-1}$ are consistent
with those of `normal' ellipticals with X-ray emission due
to a hot gaseous halo. The hardness ratio upper limit ($\rm HR <0.05$)
indicates moderate photoelectric absorption $\la6\times10^{21}\,\rm
cm^{-2}$ ($\Gamma=1.7$). We note that the optical and X-ray properties
of this source are reminiscent to those of X-ray bright optically
inactive galaxies  (Griffiths et
al. 1995; Fiore et al. 2000; Comastri et al. 2002). Apart from the
prsesence hot interstellar medium, other scenarios for the origin of
the X-ray emission of TGN\,336Z130  include (i) hot gas from a galaxy
group, (ii) radiatively inefficient Advection Dominated Accretion Flow
(ADAF), (iii) a LINER and (iv) a heavily obscured AGN. The X-ray
source associated with TGN\,336Z130 is not extended excluding the
possibility  of group emission. An ADAF or a LINER with the expected
optical emission lines diluted  by the host  galaxy stellar light can
explain the observed X-ray emission. This is particularly true in the
case of spectra obtained through wide slits/fibres (like 2dF) that are
contaminated by non-nuclear emission from the host galaxy (e.g. Ho,
Filippenko \& Sargent 1997; Moran et al. 2002; Severgnini et
al. 2003). Comastri et al. (2002) in their multiwavelength study of
FIORE-P3, the prototype of X-ray bright optically inactive systems,
favor the heavily obscured Compton thick AGN scenario with  the
observed X-ray emission arising from a scattered nuclear
component. However, for TGN\,336Z130 this possibility is inconsistent
with the  observed soft X-ray spectral properties of this sources. We
adopt a conservative approach and classify  TGN\,336Z130 an AGN. The
low X-ray luminosity suggests a LLAGN. We note however, that this may
also be a  `normal' galaxy with the X-ray emission arising from the
hot interstellar medium.  

{\bf TGN\,336Z208:} The optical counterpart of this X-ray source has
extended optical light profile. Inspection of the  SDSS image in
Figure \ref{fig_ngp} suggests spiral galaxy morphology. The 2dFGRS
spectrum reveals a broad-line AGN at $z=0.0736$. The X-ray luminosity
of this source $L_X(0.5-8)=1.8\times10^{42}\rm \, erg\, s^{-1}$
although elevated compared to `normal' galaxies is lower than that of
distant QSOs. The X-ray--to--optical flux ratio of 
$\approx-1$ place this galaxy in the border-line between 
AGNs and `normal' galaxies. The estimated hardness ratio, $\rm
HR=-0.51\pm0.15$, suggests a soft X-ray spectrum consistent with
the observed broad optical lines.  

{\bf TGN\,336Z232:} The 2dFGRS optical spectrum of this system
suggest an early type spiral with both emission ($\rm H\alpha$, $\rm
[N\,II]\,6583\AA$, $\rm [O\,II]\,3727\AA$) and absorption (H+K, $\rm
H\beta$, $\rm H\gamma$, $\rm H\delta$, $\rm NaD\,5893\AA$) lines. For
the $\rm [O\,II]\,3727\AA$ and the $\rm H\delta$ lines we estimate
equivalent widths of the $\rm EW_{[O\,II]}\approx12$
(emission) and $\rm EW_{H\delta}\approx-8\AA$ (absorption)
respectively. On the basis of the Dressler et al. (1999)
classification scheme TGN\,336Z232 is e(a) type galaxy. Poggianti et
al. (1999) discuss that the optical spectral properties of this class
of galaxies can be explained by a starburst event followed by a period
of much low star-formation activity. They also argue that in the local 
Universe e(a) type galaxies are frequently associated with
dusty interacting/merging systems. Assuming that the $\rm
[O\,II]\,3727\AA$ line of  TGN\,336Z232 is due to star-formation, the
observed X-ray emission may arise from a starburst event in the recent
past.  The  X-ray--to--optical flux ratio of $-1.5$ and the X-ray
luminosity,  $L_X(0.5-8)\approx8\times10^{41}\rm\,erg\,s^{-1}$, 
cannot discriminate between an AGN or X-ray properties dominated by
stellar processes. The hardness ratio  $3\sigma$ upper limit ($\rm HR
<0.44$) corresponds to photoelectric absorption
$\la1.5\times10^{22}\,\rm cm^{-2}$  ($\Gamma=1.7$) and cannot strongly
constrain the X-ray spectral  properties or the nature of this source.  
We adopt a conservative approach and classify this source an AGN. The
low X-ray luminosity suggest LLAGN.   

{\bf UM\,603:} This narrow emission line galaxy is classified as H\,II on
the basis of its optical spectral properties (Terlevich et
al. 1991). Its X-ray luminosity
$L_X(0.5-8)=1.5\times10^{41}\rm\,erg\,s^{-1}$ and
X-ray--to--optical flux ratio $\approx-2.1$ also suggest `normal'
galaxy. Moreover, this source is not detected in the hard-band 
suggesting a soft X-ray spectrum, $\rm HR < -0.14$, consistent
with stellar origin of the X-ray emission. 

{\bf TGN\,336Z243:} In the SDSS optical image this galaxy appears to be
interacting with a nearby smaller system. The X-ray centroid is offset
by 3\,arcsec from the optical centre of the galaxy. This is a narrow
emission line system at $z=0.0717$ showing $\rm
[O\,III]\,4959+5007\AA$ doublet, $\rm H\alpha$ and
$\rm [N\,II]\,6583\,\AA$. Although the signal--to--noise ratio of the
spectrum is low, $\rm H\beta$ is not visible in emission. This coupled
with the strong $\rm [O\,III]\,5007\AA$ feature may suggest a Seyfert 2
type system. A hardness ratio has not been estimated for this source
because it is not detected in either  the 0.5-2\,keV or the 2-8\,keV
images to the formal {\sc ewavelet}  detection threshold of $5\sigma$.   
Moreover, the low  X-ray luminosity
$L_X(0.5-8)=1.9\times10^{41}\rm\,erg\,s^{-1}$ and 
X-ray--to--optical flux ratio $\approx-2.0$ are consistent with a
`normal' galaxy powered by stellar processes.

{\bf B005615.26-275548.8:} The X-ray centroid lies
5\,arcsec off the optical source centre. The probability of chance
coincidence is $1.5\%$. The  counterpart of the X-ray source (assuming
it is the correct one) is optically extended   
and has no optical spectroscopic information. The optical--to--X-ray
flux ratio of $\approx-1.6$ place  this source in the `normal' galaxy
regime.  The hardness ratio upper limit ($\rm HR <0.40$) indicates
photoelectric absorption $\la1.5\times10^{22}\,\rm cm^{-2}$
($\Gamma=1.7$) and cannot strongly  constrain the X-ray spectral
properties or the nature of this source. On the basis of the above
evidence alone we cannot conclude on the nature of this source.  

{\bf  Other sources:} A number sources (ID numbers in Table
\ref{tbl2}: \#3, 12, 20, 23, 26) 
in the present sample (have no optical spectral
information. However, they share common properties: unresolved optical
light profile, soft X-ray spectra and  low X-ray--to--optical flux
ratio that strongly suggest Galactic stars.  Sources in the North
Galactic Pole region (\#3, \#12) also have colour information from the
SDSS. Source \#3 however, is too bright ($B\approx13.5$\,mag) and
saturated on the SDSS images. This does not allow meaningful colours to
be estimated.  Source \#12 has $g-r$, $u-g$ colours of 0.5 and 1.5
respectively, typical of stellar colours (e.g. Stoughton et
al. 2002, their Fig. 13).


\begin{thebibliography}{} 

{\bibitem{1} Alexander D. M., Aussel H., Bauer F. E., Brandt W. N.,
Hornschemeier A. E., Vignali C., Garmire G. P., Schneider D. P.,
2002, ApJ, 568, L85.}


{\bibitem{3}Barger A. J., Cowie L. L., Brandt W. N., Capak P.,
Garmire G. P., Hornschemeier A. E., Steffen A. T., Wehner E. H.,
2002, AJ, 124, 1839.}

{\bibitem{4} Bauer F. E., Alexander D. M., Brandt W. N.,
Hornschemeier A. E., Vignali C., Garmire G. P., Schneider D. P.,
2002, AJ, 124, 2351}

{\bibitem{5}  Bertin E., Arnouts S., 1996, A\&AS, 117, 393}


{\bibitem{7} Brandt W. N., Hornschemeier A. E., Schneider D. P.,
Alexander D. M., Bauer F. E., Garmire G. P., Vignali C., 2001a, ApJ,
558, 5.}

{\bibitem{8} Brandt W. N., Hornschemeier A. E., Alexander D. M.,
Garmire G. P., Schneider D. P., Broos P. S., Townsley L. K., Bautz
M. W., Feigelson E. D., Griffiths R. E., 2001b, AJ, 122, 1.}


{\bibitem{9} Brandt W. N., 2001c, AJ, 122, 2810.}


{\bibitem{11} Croom S. M., Smith R. J., Boyle B. J., Shanks T., Loaring
N. S., Miller L., Lewis I. J., 2001, MNRAS, 322L, 29}

{\bibitem{12} Colless M., Dalton G., Maddox S. et al.,  2001, MNRAS,
328, 1039.}

{\bibitem{13} Comastri A., et al. 2002, ApJ, 571, 771.}



{\bibitem{16} Downes A. J. B., Peacock J. A., Savage A., Carrie D. R.,
1986, MNRAS, 218, 31}

{\bibitem{16a} Dressler A., Smail I., Poggianti B. M., Butcher H.,
Couch W. J., Ellis R. S., Oemler A. Jr.,  1999, ApJS, 122, 51}


{\bibitem{16b} Elvis M., Soltan A., Keel W., 1984, ApJ, 283, 479}


{\bibitem{19}  Fiore F., La Franca F., Vignali C., Comastri A., Matt
G., Perola G. C., Cappi M., Elvis M.,  Nicastro F.,  2000, NewA, 5, 143}



{\bibitem{22} Fukugita M., Ichikawa T., Gunn J. E., Doi M.,
Shimasaku K., Schneider D. P., 1996, AJ, 111, 1748} 


{\bibitem{25} Georgakakis A, Georgantopoulos I., Stewart G. C., Shanks
T., Boyle B. J., 2003a, MNRAS, in press, astro-ph/0305278}

{\bibitem{26} Georgakakis A., Hopkins A. M., Sullivan M., Afonso J.,
Georgantopoulos I., Mobasher B.,  Cram L. E., 2003b, MNRAS, submitted}  



{\bibitem{27} Georgantopoulos I., Basilakos S., Plionis M., 1999,
MNRAS, 305, L31.} 
{\bibitem{28} Georgantopoulos I., Stewart G. C., Shanks T., Boyle B. J., Griffiths R. E., 1996, MNRAS, 280, 276.}








{\bibitem{36}Griffiths R. E.,  Georgantopoulos I., Boyle B. J.,
Stewart G. C., Shanks T., Della Ceca R., 1995, MNRAS, 275, 77.}


{\bibitem{37}Griffiths R. E.,  Della Ceca R., Georgantopoulos I., Boyle B. J.,
Stewart G. C., Shanks T., Fruscione A., 1996, MNRAS, 281, 71.}



{\bibitem{39}  Hasinger G., Altieri B., Arnaud, M. et al., 2001,
A\&A, 365, L45.}



{\bibitem{42} Ho L. C., Filippenko A. V., Sargent W., 1995, ApJS, 98,
477.}  

{\bibitem{43} Ho L. C., Filippenko A. V., Sargent W., 1997, ApJS, 112,
315.} 

{\bibitem{44} Hornschemeier A. E., Brandt W. N., Alexander D. M., Bauer
F. E., Garmire G. P., Schneider D. P., Bautz M. W., Chartas G., 2002a,
ApJ, 568, 82.}

{\bibitem{45} Hornschemeier A. E., Bauer F. E., Alexander D. M., Brandt
W. N., Sargent W. L. W., Vignali C., Garmire G. P., Schneider D. P.,
2003, AJ, accepted, astro-ph/0305086.}

{\bibitem{47} Kim D. W., Fabbiano G., Trinchieri G., 1992, ApJ, 393, 134.}

{\bibitem{48}  Landolt A. U., 1992, AJ, 104, 340}

{\bibitem{49} Lehmann I., et al.,  2001, A\&A, 371, 833.}
{\bibitem{50} Lilly S. J., Le Fevre O., Hammer F., Crampton D., 1996,
ApJ, 460, L1.}

{\bibitem{51} Maddox S. J., Sutherland W. J., Efstathiou G., Loveday J.,
Peterson B. A.,  1990, MNRAS, 247L, 1}


{\bibitem{53}  McHardy I. M., et al., 1998, MNRAS, 295, 641}

{\bibitem{54} McHardy I. M., Gunn K. F., Newsam A. M.,  Mason K. O.,
Page M. J., Takata T., Sekiguchi K., Sasseen T., Cordova F., Jones
L. R., Loaring N., 2003, MNRAS, in press, astro-ph/0302553}


{\bibitem{55} Metcalfe N., Fong R., Shanks T., 1995, MNRAS, 274, 769}

{\bibitem{55a}
Miyaji T., Griffiths R. E., 2002, ApJ, 564L, 5}






{\bibitem{57} Moran E. C., Lehnert M. D., Helfand D. J., 1999, ApJ, 526, 649}

{\bibitem{57a} Moran E., Filippenko A. V., Chornock R.,  2002, ApJ,
579L, 71}



{\bibitem{60} O'Sullivan E., Forbes D. A., Ponman T. J., 2001, MNRAS,
328, 461.}


{\bibitem{60a} Poggianti B. M., Smail I., Dressler A., Couch W. J.,
Barger A. J., Butcher H., Ellis R. S., Oemler A. Jr.,  1999, ApJ, 518,
576}




{\bibitem{66} Roche N., Griffits R. E., Della Ceca R., Shanks T.,
Boyle B. J., Georgantopoulos I., Stewart G. C., 1996, MNRAS, 282, 820} 

\bibitem{66a} Severgnini P. et al., 2003, A\&A, 406, 483


{\bibitem{70} Stocke J. T., Morris S. L., Gioia I. M., Maccacaro T.,
Schild R., Wolter A., Fleming T. A., Henry J. P., 1991, ApJS, 76, 813}   


{\bibitem{72} Stoughton C., et al.,  2002, AJ, 123, 485.}

{\bibitem{73} Str\"uder L., Briel U., Dennerl K., et al. 2001, A\&A, 365, L18.}

{\bibitem{76} Turner M. J. L., Abbey A., Arnaud M., et al., 2001, A\&A,
365, L27.}

{\bibitem{77} York D. G., et al., 2000, AJ, 120, 1579.}

\end{thebibliography}
\end{document}